\documentclass[%
 reprint,
nofootinbib,
 amsmath,amssymb,
 aps,
prd,
noeprint,
nolongbibliography,
]{revtex4-2}

\usepackage{graphicx}
\usepackage{dcolumn}
\usepackage{bm}
\usepackage{textcomp}
\usepackage{hyperref}
\usepackage{multirow}
\usepackage{color,ulem}

\newcommand{\aap}{Astron. Astrophys.}
\newcommand{\aj}{Astron. J.}
\newcommand{\apjl}{\apj\ Lett.}
\newcommand{\mnras}{Mon. Not. R. Astron. Soc.}
\newcommand{\nus}{\nu_{\rm s}}
\newcommand{\nugw}{\nu_{\rm gw}}
\newcommand{\tgw}{\tau_{\rm gw}}
\newcommand{\css}{c_{\rm s}^2}

\begin{document}

\preprint{APS/123-QED}

\title{Gravitational waves from glitch-induced f-mode oscillations in quark and neutron stars}

\author{Oliver H. Wilson}
 \email{owilson@haverford.edu}

\author{Wynn C. G. Ho}
 \email{Corresponding author; who@haverford.edu}

 \affiliation{ 
Department of Physics and Astronomy,
Haverford College,
370 Lancaster Avenue,
Haverford, Pennsylvania 19041, USA
}%

\date{\today}

\begin{abstract}
Matter in compact stars is dense enough that transient events within the star could have sufficiently high energies to produce detectable gravitational waves (GWs). These GWs could be used to constrain the equation of state (EoS) for matter in the star and could reveal that there is more than one type of EoS at play in the population, implying that multiple types of compact stars exist. One of these types could be quark stars, composed almost entirely of stable quark matter, and observing GWs is a way to test for the strange matter EoS. Here we explore the possibility that, if fundamental (f-) mode oscillations in pulsars are induced by a pulsar glitch, then these oscillations might produce detectable GWs. We use the existing population of pulsars and their glitches, as well as a much larger synthesized population, along with 15 EoSs (8 for neutron stars and 7 for quark stars) to generate frequencies, damping times, and GW strengths for each. We find that of the EoSs examined, all quark star EoSs produce narrower distributions of f-mode frequency than neutron star EoSs.  This result, along with other elements of the data, could be used to differentiate between GWs (or other signals from f-modes) originating from neutron stars and quark stars and thus could confirm the existence of quark stars. We also find that GW astronomy is a potentially viable method for detecting a larger population of pulsars which are not observable electromagnetically and that future GW observatories have the possibility to greatly expand this capability. 
\end{abstract}

\maketitle


\section{\label{sec:level1}Introduction}
The field of gravitational wave (GW) astronomy formally began less than a decade ago with the detection of GW150914 originating from a binary black hole merger by the Laser Inferometer Gravitational-wave Observatory (LIGO) \cite{GW150914}. Since then, GW astronomy has focused on very high energy, often extragalactic, events. These events include binary black hole and neutron star (NS) mergers, supernovae, and most recently, the supermassive black hole mergers and possibility of a cosmic gravitational wave background detected by the NanoGrav collaboration \cite{Agazie_2023}. Despite this focus on high-distance and high-energy events, the LIGO, Virgo, and KAGRA (LVK) GW observatories are capable of detecting lower-energy but much lower-distance events as well (see, e.g., \cite{Riles_2023,Wette_2023}). As these detectors are upgraded and next generation detectors are built, they may begin to detect signals of this nature. Possible sources of this category of signals include pulsar glitches \cite{Yim_2023,Haskell_2024} and oscillations in NSs, particularly oscillations of the fundamental mode or f-mode \cite{Andersson_2021}. GWs from the latter could be used to probe the stellar mass $M$ and radius $R$, which would allow the constraining of the Equation of State (EoS) for the matter in NSs \cite{Andersson1998TowardsGW,Andersson_2001}. F-mode oscillations can occur as the result of NS births and mergers, as well as other causes such as  pulsar glitches \cite{Prix_2011,Keer_2015,Ho_2020,Yim_2023} (see \cite{Antonelli_2022,Antonopoulou_2022,Zhou_2022}, for review of pulsar glitches). \cite{Ho_2020} explores the characteristics of GWs from glitch-induced f-mode oscillations in NSs using the existing population of known pulsar glitches but was limited in scope.  Firstly, only one EoS for NS matter was considered, i.e., BSk24 \cite{Pearson_2018}. Each EoS has a different relationship between properties of the star, such as its mass, and the f-mode frequency $\nugw$ and damping time $\tgw$, so by exploring multiple EoSs we get a more comprehensive understanding of the characteristics of the distributions of these values in the model. We note that another approach to probing nuclear properties, such as  nucleon effective mass, incompressibility, and symmetry energy, is to consider a single EoS and determine constraints on these properties based on GW detection of glitch-induced f-modes (see, e.g., \cite{Pradhan_2023}). A second limitation of the study by \cite{Ho_2020} is that the known population of pulsar glitches is relatively small, with only 650 glitches across 200 pulsars. By using a synthesized population of glitches alongside the known population, we can see the effects of a larger number of pulsars and correct for some biases and statistical anomalies in the real data. Finally, \cite{Ho_2020} only considers NSs, while there exists another hypothesized object that could be responsible for at least some of the pulsar population, i.e., quark stars. 
Here we follow-up \cite{Ho_2020} by considering the above limitations; see \cite{Ho_2020}, and references therein, for more discussion of the context of our study.

Quark stars (QSs) are a hypothesized type of compact star and are similar to NSs in many ways. 
Quarks are deconfined throughout most of the star rather than being bound in nucleons
\cite{Weber_2005}. The most likely form of quark matter, if it exists, is strange matter,
which is created when the high densities in collapsed stars make it energetically favorable for down quarks to become strange quarks.
The theoretical basis of stars composed entirely or almost entirely of strange matter is linked with the strange matter hypothesis \cite{Weber_2005}. The strange matter hypothesis proposes that the current form of baryonic matter, dominated by protons and neutrons composed of up and down quarks, is only metastable, and that the most stable form of matter would be charge-neutral strange matter composed of baryons with up, down, and strange quarks.
QSs would likely only represent some fraction of the population of pulsars because some pulsars have been observed to be very consistent with NS models.
Therefore, if both NSs and QSs exist and glitches are capable of inducing f-mode oscillations, observation of GWs from glitches could be a promising way to confirm the existence of QSs and provide evidence in favor of the strange matter hypothesis. Furthermore, even if glitches do not produce detectiable GW f-modes, they provide a convenient way to explore NS and QS f-modes in general (see, e.g., \cite{Kojima_2002,Sotani_2003}, for complementary ways). These results can then be applied to all f-modes regardless of origin or method of detection. In this paper, we explore the characteristics of GWs from f-mode oscillations in NSs and QSs, along with ways to differentiate between the two in GW or other f-mode data. 

\section{Methodology}
\subsection{Real and Synthesized Glitch Populations}
Three populations of pulsars with corresponding glitches were generated for the purposes of this paper. Population 1 consists of 650 recorded glitches across 195 real pulsars. We took the 650 glitches from the JBCA Glitch Catalogue\footnote{\href{https://www.jb.man.ac.uk/pulsar/glitches.html}{https://www.jb.man.ac.uk/pulsar/glitches.html}; accessed on May 15 2023}
\cite{Espinoza_2011,Basu_2022} and supplemented these data with information about their respective pulsars from the ATNF Pulsar Catalogue (\cite{Manchester_2005}; version 1.69). From these two catalogs, we took the observed values for spin frequency $\nus$, glitch size $\Delta\nus$, and distance $d$. We randomly assigned masses to each of the pulsars according to a Gaussian distribution with a mean \(\mu_M=1.4\,M_\odot\) and width \(\sigma_M=0.15\,M_\odot\), which is the same as used in \cite{Ho_2020}. Population 2 is identical to Population 1, except its masses were assigned to pulsars according to a uniform distribution from \(1\,M_\odot\) to \(2\,M_\odot\). Population 2 was included to study the effects of a broader distribution of mass, and thus broader f-mode frequency and damping time distributions via $\nugw(M)$ and $\tgw(M)$ (see below).
For Populations 1 and 2, the fact that every pulsar, even those that glitched multiple times, was assigned a particular mass means that a handful of frequently-glitching pulsars stand out in some of the relevant figures, especially the nearby Vela pulsar. In contrast, Population 3 has no pulsars with multiple glitches because it consists of 10000 randomly generated pulsars and glitches, with each glitch having its own corresponding pulsar. For Population 3, we randomly generated mass, spin frequency, glitch size, and distance. Mass was generated according to the same Gaussian distribution as in Population 1. Spin frequency was generated according to a truncated Gaussian distribution for spin period ($P=1/\nus$), with mean \(\mu_P=300\,\mbox{ms}\) and width \(\sigma_P=150\,\mbox{ms}\) from the values found in \cite{Faucher_Giguere_2006}. Glitch size was generated using the bimodal Gaussian distributions in \cite{Fuentes_2017}.  Using the Galactic distribution of pulsars and the location of the Sun in \cite{Faucher_Giguere_2006}, a Gaussian distribution was generated for the distance of pulsars from the Sun, with mean \(\mu_d=11.8\,\mbox{kpc}\), width \(\sigma_d=5.8\,\mbox{kpc}\), and minimum distance of 100 pc. In Population 3, rather than precisely modeling the real population of pulsars, the goal is instead to generate plausible results, avoiding complete reliance on the current population of known glitching pulsars. This is because the known population has well-known observational biases, the simplest of which is distance-related effects such as brightness and obscuration.

Spin frequency, glitch size, distance, and glitch energy ($E_{\rm glitch}=4\pi^2I\nus\Delta\nus$ for simplicity, where \(I\sim10^{45}\ \mbox{g cm}^2 \) is the stellar moment of inertia) are all either observed (synthesized) values in Populations 1 and 2 (Population 3) or are calculated from these values independently of the EoSs and assigned masses. Therefore, Populations 1 and 2 are identical with respect to these values and are represented together in the figures in this section, while Population 3 is represented separately. 

\begin{figure}
    \centering
    \includegraphics[width=0.48\textwidth]{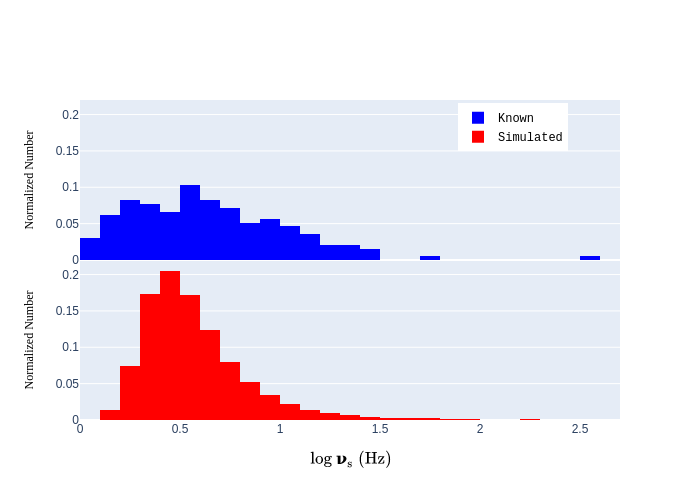}
    \caption{Normalized distributions of spin frequency \(\nus\) from 195 observed pulsars (top) and from $10^4$ synthesized pulsars (bottom).} 
    \label{Fig. 1}
\end{figure}

\begin{figure}
    \centering
    \includegraphics[width=0.48\textwidth]{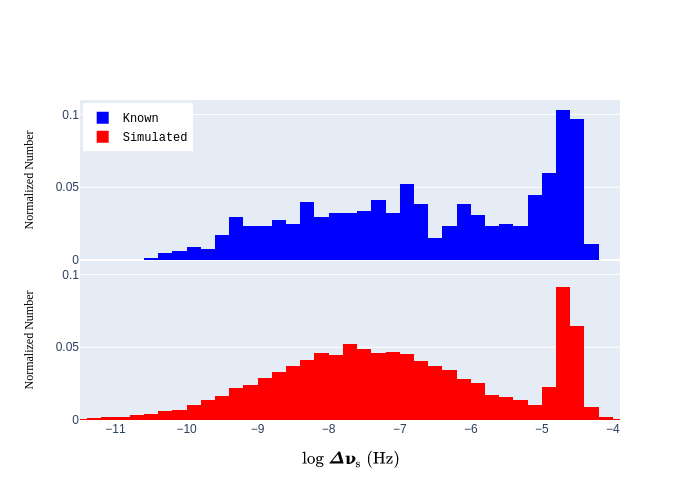}
    \caption{Normalized distributions of glitch size \(\Delta\nus\) from 195 observed pulsars (top) and from $10^4$ synthesized pulsars (bottom).}
    \label{Fig. 2}
\end{figure}

\begin{figure}
    \centering
    \includegraphics[width=0.48\textwidth]{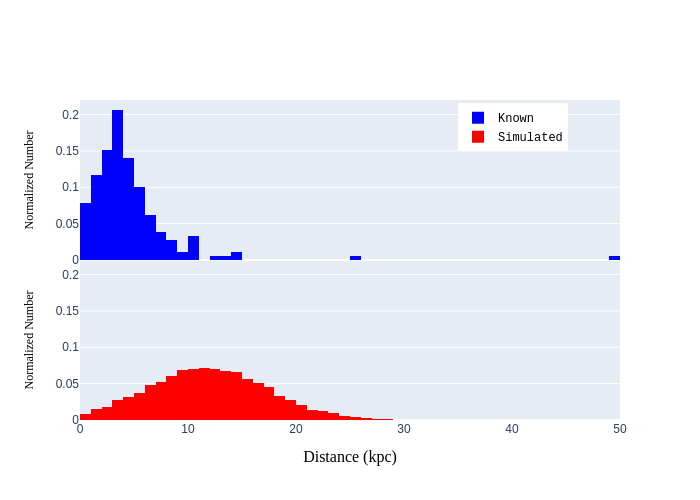}
    \caption{Normalized distributions of distance from the Sun from 195 observed pulsars (top) and from $10^4$ synthesized pulsars (bottom).}
    \label{Fig. 3}
\end{figure}

\begin{figure}
    \centering
    \includegraphics[width=0.48\textwidth]{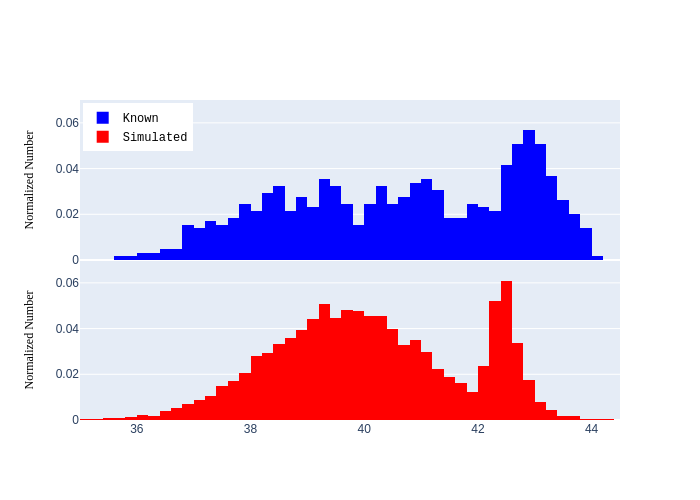}
    \caption{Normalized distributions of glitch energy \(E_{\rm glitch}\) from 195 observed pulsars (top) and from $10^4$ synthesized pulsars (bottom).}
    \label{Fig. 4}
\end{figure}

In Fig.~\ref{Fig. 1}, we can see an approximately normal distribution of spin frequency in the observed glitching pulsar population (Pops. 1 and 2) and a similar distribution in the synthesized population (Pop. 3). Figure~\ref{Fig. 2} displays a bimodal or trimodal distribution of glitch sizes in the known glitch population, along with the synthesized bimodal population. Figure~\ref{Fig. 3} shows the distribution of distance from the Sun for the population of known glitching pulsars, with a clear observational bias towards those nearby.  Meanwhile, the synthesized population is a simple Gaussian and is meant to approximate the Galactic population of pulsars without this observational bias. By using this latter distance distribution for Population 3, we can make more general predictions about the possibility of detecting glitches from the Galactic population of pulsars by GW observatories, including those pulsars which have yet to be detected using electromagnetic telescopes and non-pulsar rotating NSs and QSs which glitch (non-pulsar in the sense that their pulsing emissions do not point towards Earth). Finally, Fig.~\ref{Fig. 4} demonstrates the energy of glitches in both groups, with clear similarities to the distribution of glitch sizes, as would be expected. Population 3 has fewer very high-energy glitches because Populations 1 and 2 include PSR~J0537$-$6910, also known as the Big Glitcher, which has more than 50 large glitches in the population of known glitches (e.g., \cite{Antonopoulou_2018,Ho_2020b}), which combined with its high spin frequency creates a large group of unusually high-energy glitches in the observed data.  Note however that PSR~J0537$-$6910 is in the Large Magellanic Cloud at a distance of 50~kpc.  Therefore, Figs.~\ref{Fig. 1}--\ref{Fig. 4} show that Population 3 is an appropriate approximation of the observed population of pulsars and pulsar glitches.

\subsection{$\nugw$ and $\tgw$ as Functions of M and R}
F-mode frequency and damping time are generally functions of stellar density ($\propto M/R^3$) and compactness ($\propto M/R$), respectively \cite{Andersson1998TowardsGW}.
Radius itself is a function of mass for a given EoS, so it is possible to find f-mode frequency $\nugw(M)$ and damping time $\tgw(M)$ from our randomly generated mass values.
Because the EoS for NSs and QSs is not known, we need to find a different $\nugw(M)$ and $\tgw(M)$ for each EoS, either by using $\nugw(M)$ and $\tgw(M)$ calculated directly from an EoS and given in the literature or by determining $R(M)$ and then using general relations for $\nugw(M)$ and $\tgw(M)$. 

\subsection{Equations of State}
We considered a number of EoSs to generate the data we require.
In total, we used 15 EoSs, 8 of them for NSs and 7 for QSs. The $\nugw (M)$ and $\tgw(M)$ relations for the ZL NS EoS was taken directly from \cite{Zhao_Lattimer_2022} (see also \cite{Drischler_2021}). The \(R(M)\) relations for BSk24 and APR NS EoSs were taken from \cite{Pearson_2018} and \cite{APR_1998} respectively, and put through the \(\nugw(M//R^3)\) and \(\tgw(M/R)\) relations in \cite{Doneva_2013}.
We took the \(R(M)\) functions for the SLy4, SHT, BBB2, GNH3, and LS220 NS EoSs from \cite{Chirenti_2015} and used the \(\nugw(M/R^3)\) and \(\tgw(M/R)\) functions therein. We used the \(R(M)\) relation for QS EoS 1 (hereafter CMOT1) from \cite{CeliMarcosOsvaldo2022OMCS}
with the BFG, CFL, and magnetized CFL fits relating f-mode frequency to density and damping time to compactness from the same paper. \cite{CeliMarcosOsvaldo2022OMCS} also included a slightly different version of the same EoS 1 for magnetars, which we used with the magnetized CFL fit (hereafter CMOT1m). Finally, the $\nugw(M)$ and $\tgw(M)$ relations in \cite{Zhao_Lattimer_2022} for three parameterizations of a QS EoS were used, specifically for \(M_{\rm max}=2.2\,M_{\odot}\) and \(\css=1/3\), \(M_{\rm max}=2.2\,M_{\odot}\) and \(\css=1/2\), and \(M_{\rm max}=2.2\,M_{\odot}\) and \(\css=1\). Because  \(M_{\rm max}=2.2\,M_{\odot}\) for all three ZL QS EoSs, hereafter these EoSs will be distinguished only by their \(\css\) value.

\subsection{Glitch Energy, GW Strain, and GW Detector Sensitivity} \label{sec:gwglitch}
To generate values for glitch energy and GW strain for each glitch, the relations in \cite{Ho_2020} are used. First, an estimated glitch energy is expressed as a function of spin frequency and glitch size:
\begin{equation}
    E_{\rm glitch}=3.95\times10^{40}\,\mbox{erg}\left(\frac{\nus}{10\,\mbox{Hz}}\right)\left(\frac{\Delta\nus}{10^{-7}\,\mbox{Hz}}\right).\label{eq:1}
\end{equation}
Second, we assume this glitch energy powers GWs, with a peak GW amplitude \(h_0\) is expressed as a function of distance, spin frequency, glitch size, GW frequency, and GW damping time:
\begin{eqnarray}
h_0&=&7.21\times10^{-24} \left(\frac{1\,\mbox{kpc}}{d}\right)\left(\frac{\nus}{10\,\mbox{Hz}}\right)^{1/2}\left(\frac{\Delta\nus}{10^{-7}\, \mbox{Hz}}\right)^{1/2} \nonumber \\
& & \times\left(\frac{1\,\mbox{kHz}}{\nugw}\right)\left(\frac{0.1\,\mbox{s}}{\tgw}\right)^{1/2}.
\label{eq:2}        
\end{eqnarray}
The calculated peak amplitude is used with damping time to calculate GW strain  ($=h_0\sqrt{\tgw}$). 
It is important to note that the relations from \cite{Ho_2020} assume 100\% efficiency in transferring glitch energy into GW energy. Since the actual efficiency is unknown, the GW strains modeled in this study are by definition overestimated to some degree, and possibly to a significant degree, and thus are representative upper limits. Therefore, while f-mode oscillations will generate \textit{some} GW signal, it is possible that the signal would be too weak to measure, either with current detectors or any detectors in the near future.\\

Sensitivity curves for the following GW detectors are included in some of the figures and analysis: aLIGO (the current version of LIGO) \cite{aLIGO}, A+ (a future upgrade to LIGO) \cite{a+LIGO}, and the planned Einstein Telescope (ET) \cite{hild2008pushing} and Cosmic Explorer (CE) \cite{reitze2019cosmic}. 

\section{Results}
GW frequency $\nugw$ and damping time $\tgw$ are calculated for every
glitch in the 3 populations and for each of the 15 EoSs.
These are then used to calculate the GW strain
[$=h_0\sqrt{\tgw}$; see equation~(\ref{eq:2})]
and signal-to-noise ratio (SNR), where SNR$=h_0\sqrt{\tgw/2S_h}$ and
$\sqrt{S_h}$ is the spectral noise density of GW detectors obtained from
\cite{aLIGO,a+LIGO,hild2008pushing,reitze2019cosmic}.
Figures showing $\nugw$, $\tgw$, and GW strain for all 3 populations
and all 15 EoSs are in Appendix~\ref{sec:appendix}, while only a subset of these are described here.

In particular, in this section, we show results from only Population~1
(i.e., known pulsars and glitches and Gaussian mass distribution)
and using a few of the 15 EoSs;
the figures hereafter are all plotted on the same respective scales to
enable easier comparisons between various EoSs.
The few EoSs used here are selected simply because they illustrate
three possible scenarios for how NS and QS EoSs might be distinguishable
using GW detections.
These three scenarios are the following:
\begin{enumerate}
\item[A.]
Scenario 1: The NS EoS and QS EoS produce GW frequencies and damping times, as
functions of mass, that are clearly distinct from one another.
\item[B.]
Scenario 2: The NS EoS and QS EoS produce $\nugw(M)$ and $\tgw(M)$ that have some
overlap in their distributions, but there are significant differences
between their mean values.
\item[C.]
Scenario 3: The NS EoS and QS EoS produce $\nugw(M)$ and $\tgw(M)$ that have
significant overlap in their distributions.
\end{enumerate}

\subsection{Scenario~1: NS and QS EoSs with distinct $\nugw$ and $\tgw$}
\label{sec:scenario1}
\begin{figure}
    \centering
    \includegraphics[width=0.48\textwidth]{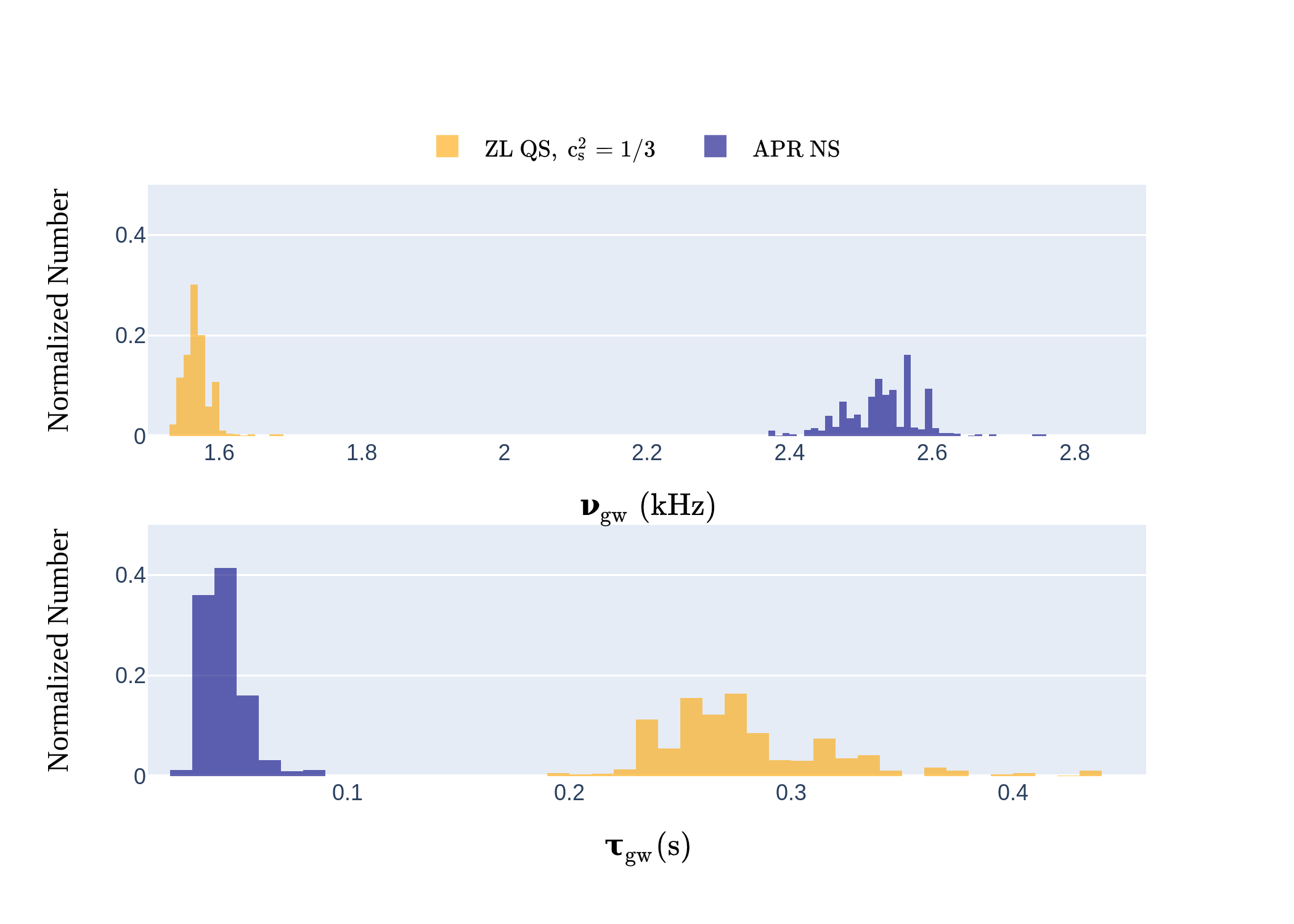}
\caption{Normalized distributions of GW frequency \(\nugw\) and damping time $\tgw$ using the ZL QS $\css=1/3$ EoS and APR NS EoS and Population 1 (known pulsars and glitches and assuming a Gaussian mass distribution).}
    \label{Fig. 5}
\end{figure}

In this scenario, we use results from the APR NS EoS and the ZL QS $\css=1/3$ EoS as examples which illustrate when NSs and QSs have distinctly different GW frequencies $\nugw$ and damping times $\tgw$. 
This can be seen from Figure~\ref{Fig. 5}, which shows normalized distributions of $\nugw$ and  $\tgw$ for each observed glitch and assuming a Gaussian mass distribution (i.e., Population~1). Not only are the NS and QS distributions different in absolute values, we see there is another very important difference between the two: the QS EoS produces a much narrower distribution of GW frequencies.  In fact, this is true across all 15 EoSs, with the widest distribution of the QS EoSs being substantially narrower than the narrowest NS EoS distribution (see below).
We note here that the effects of the different (Gaussian versus uniform) mass distributions of Populations~1 and 2 can be seen in Figure~\ref{Fig. 9}, with the wider uniform distribution of Population~2 producing visibly wider distributions of GW frequencies. Population~3 has the same mass distribution as Population~1, but the much larger number of glitches creates much smoother distributions.
For GW damping times, while Figure~\ref{Fig. 5} shows the width of the APR NS distribution is narrower than that of the ZL QS $\css=1/3$ EoS, this is not the case for all 15 EoSs we considered.

Tables~1 and 2 contain values for the widths of the frequency and damping time distributions of all 15 EoSs and for the 3 populations (see Appendix~\ref{sec:appendix} for the corresponding figures). Table~1 shows how QS EoSs have a narrower f-mode frequency distribution compared to NS EoSs, as all 7 QS EoSs have significantly lower \(\sigma_\nu\) values across the 3 populations than the narrowest NS EoS (SHT) does for the same population. Table~1 also demonstrates that, assuming a normal or near-normal distribution in the real f-mode data, sufficiently large (multiple \(\sigma\)) separations in measured frequency or damping time would indicate that two sources are unlikely to be part of the same distribution from one EoS, implying that there are two EoSs and therefore two types of star in the glitching pulsar population. On the other hand, Table~2 shows no discernible trend between NS and QS EoSs when it comes to widths. However, it is important to note that this does not mean that the real QS and NS EoSs have the same distribution (in terms of both mean and width) when it comes to damping time. In fact, they almost certainly deviate from one another if they both exist, but it is more difficult to discern differences than it is to do the same for f-mode frequencies.

\begin{table}
\begin{center}
\begin{tabular}{|c|c|c|c|}
 \hline
 \multicolumn{4}{|c|}{Table 1: \(\sigma_\nu\) (in kHz) of \(\nugw\) for each EoS and Population} \\
 \hline
QS EoS & Pop. 1 &Pop. 2&Pop. 3\\
 \hline
CMOT1 with CFL& 0.017 & 0.052 & 0.020 \\
ZL \(\css=1/3\)& 0.019 & 0.054 & 0.022 \\
CMOT1 with magnetized CFL& 0.019 & 0.056 & 0.021 \\
CMOT1m with magnetized CFL& 0.018 & 0.055 & 0.021 \\
ZL \(\css=1/2\)& 0.015 & 0.049 & 0.018 \\
CMOT1 with BFG& 0.013 & 0.039 & 0.015 \\
ZL \(\css=1\)& 0.0053 & 0.017 & 0.0060 \\
 \hline
NS EoS & & &\\
 \hline
SHT& 0.029 & 0.062 & 0.035 \\
GNH3& 0.090 & 0.23 & 0.10 \\
LS220& 0.072 & 0.19 & 0.083 \\
ZL& 0.084 & 0.21 & 0.099 \\
SLy4& 0.067 & 0.17 & 0.078 \\
BBB2& 0.091 & 0.25 & 0.11 \\
BSk24& 0.039 & 0.080 & 0.046 \\
APR& 0.054 & 0.12 & 0.063 \\
 \hline
 \end{tabular}
 \label{tab1}
\end{center}
\end{table}

\begin{table}[!]
\begin{center}
\begin{tabular}{|c|c|c|c|}
 \hline
 \multicolumn{4}{|c|}{Table 2: \(\sigma_\tau\) (in s) of \(\tgw\) for each EoS and Population} \\
 \hline
QS EoS & Pop. 1 &Pop. 2&Pop. 3\\
\hline
CMOT1 with CFL& 0.035 & 0.069 & 0.043 \\
ZL \(\css=1/3\)& 0.039 & 0.073 & 0.048 \\
CMOT1 with magnetized CFL& 0.034 & 0.068 & 0.042 \\
CMOT1m with magnetized CFL& 0.034 & 0.067 & 0.041 \\
ZL \(\css=1/2\)& 0.021 & 0.038 & 0.025 \\
CMOT1 with BFG& 0.018 & 0.034 & 0.022 \\
ZL \(\css=1\)& 0.012 & 0.018 & 0.014 \\
 \hline
NS EoS & & &\\
 \hline
SHT& 0.077 & 0.14 & 0.10 \\
GNH3& 0.078 & 0.15 & 0.098 \\
LS220& 0.047 & 0.095 & 0.060 \\
ZL& 0.032 & 0.058 & 0.041 \\
SLy4& 0.032 & 0.064 & 0.041 \\
BBB2& 0.031 & 0.054 & 0.039 \\
BSk24& 0.013 & 0.024 & 0.016 \\
APR NS& 0.0092 & 0.017 & 0.012 \\
 \hline
 \end{tabular}
 \label{tab2}
\end{center}
\end{table}

\begin{figure}
    \centering
\includegraphics[width=0.48\textwidth]{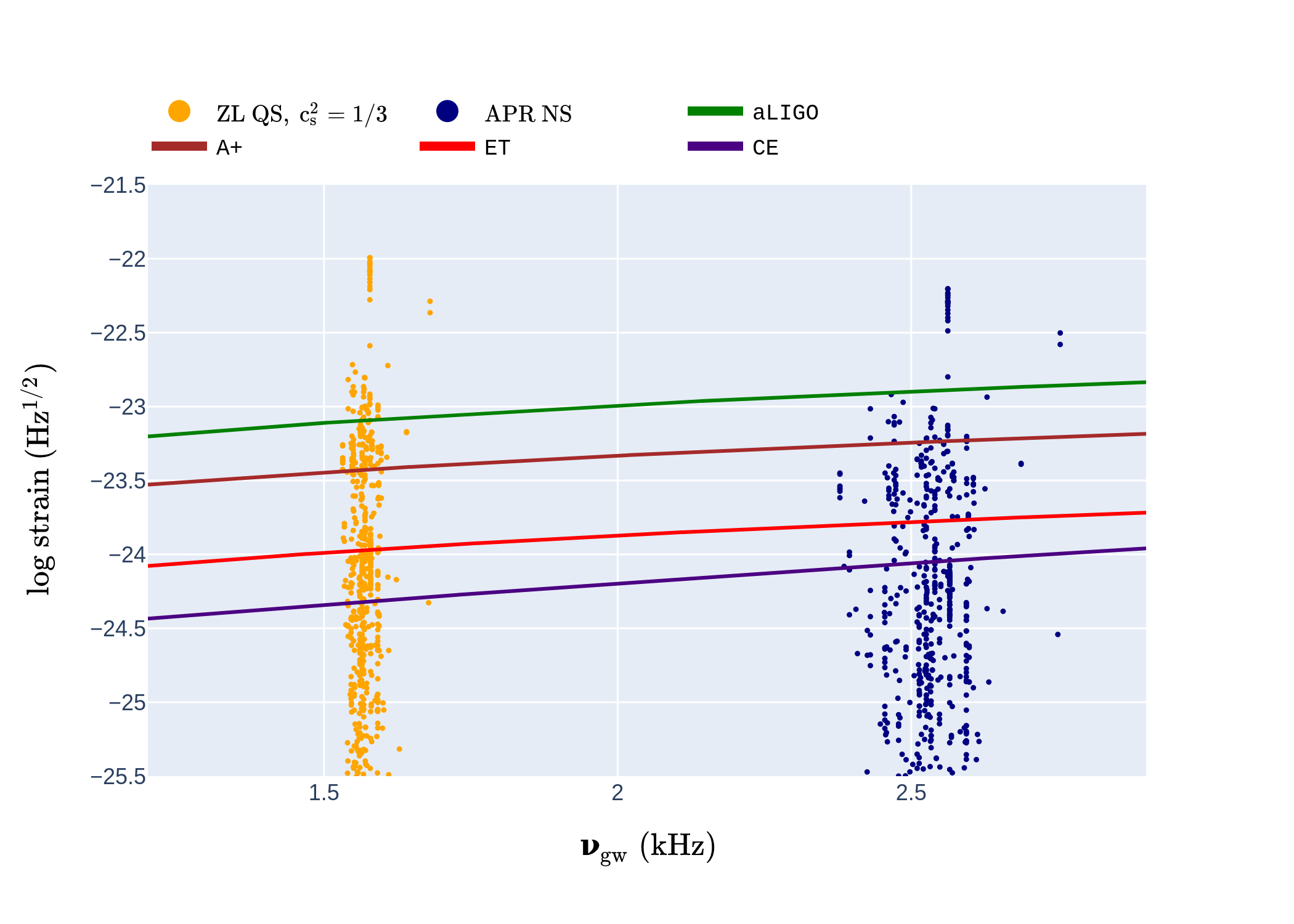}
\caption{GW strain ($h_0\sqrt{\tgw}$) from glitch-induced f-modes, calculated using the ZL QS $\css=1/3$ EoS and APR NS EoS and Population 1 (known pulsars and glitches and assuming a Gaussian mass distribution).  Solid lines indicate current and future GW detector sensitivities.
}
    \label{Fig. 6}
\end{figure}

Figure~\ref{Fig. 6} displays the GW spectra, along with sensitivity curves for current and planned GW observatories, to illustrate the portion of glitches that would be detected by a given observatory. Figure~\ref{Fig. 6} is helpful in visualizing the data that could be collected if both QSs and NSs are present in the population of glitching pulsars. For Scenario 1 with NS and QS EoSs producing very different f-mode frequencies, if there are roughly equal populations of NSs and QSs in the Galaxy, then the first few detections of GWs from glitches would likely originate from some of each kind of compact star. It would then be possible to determine that the GW spectra of the glitches are in two groups -- corresponding to two different EoSs and indicating that both NSs and QSs are responsible -- from just a handful of glitches.

\begin{figure}
    \centering
\includegraphics[width=0.48\textwidth]{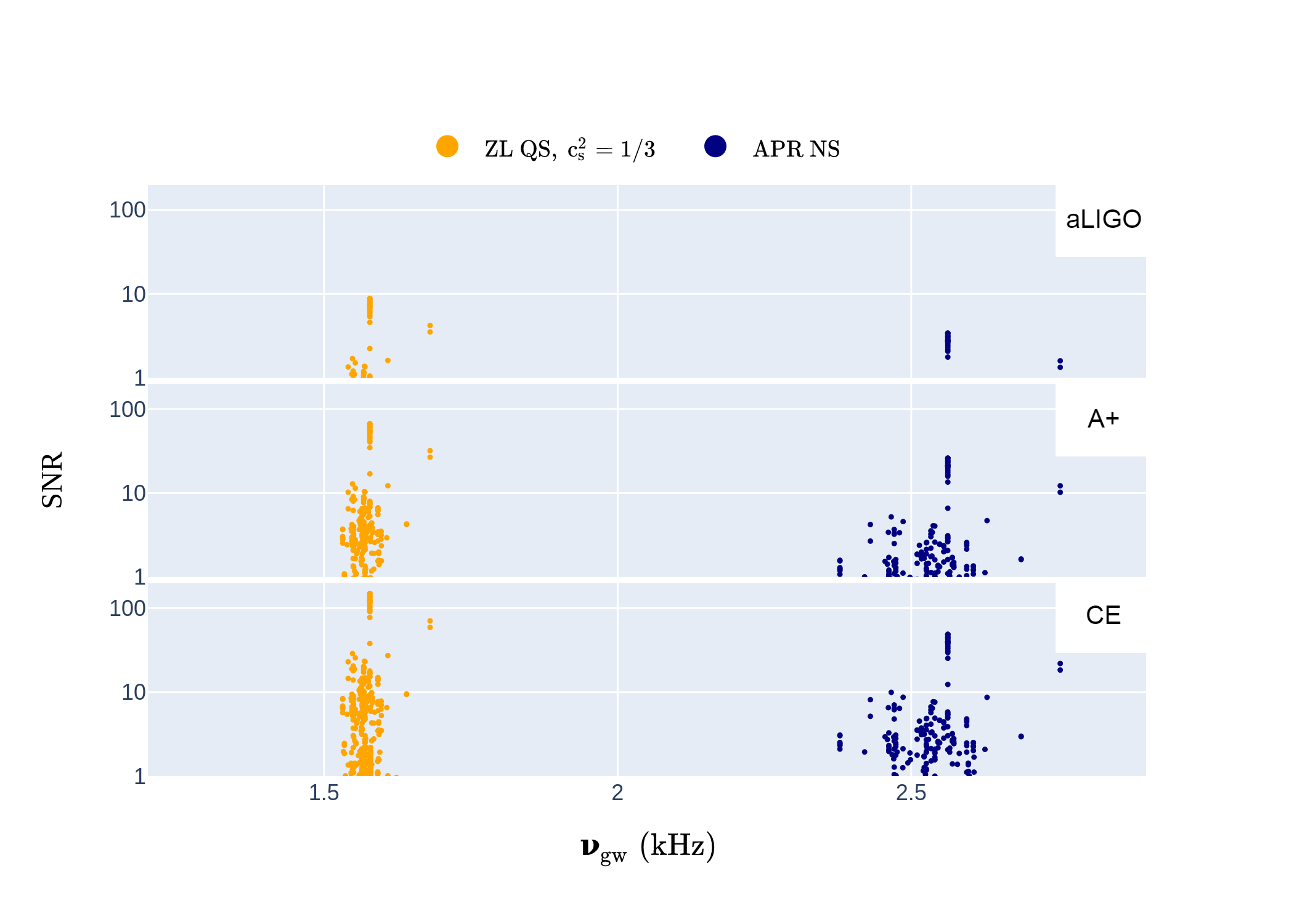}
\caption{Signal-to-noise ratio (SNR $=h_0\sqrt{\tgw/2S_h}$, where $\sqrt{S_h}$ is the spectral noise density of aLIGO, A+, or CE) from glitch-induced f-modes, calculated using the ZL QS $\css=1/3$ EoS and APR NS EoS and Population 1 (known pulsars and glitches and assuming a Gaussian mass distribution).} 
    \label{Fig. 8}
\end{figure}

Figure~\ref{Fig. 8} shows the SNRs for three GW detectors: aLIGO, A+, and CE. The peaks seen here are due to the Vela pulsar because of its close distance and the large size of its glitches.  Vela's prominence suggests it is the most likely candidate among known pulsars for a first detection of one of its glitches by aLIGO. Vela undergoes a glitch around every 18 months.  The current LIGO/Virgo/KAGRA observing run (O4) is slated to last 20 months starting from May 2023, so it is  possible that a GW originating from Vela could be detected during O4 if the responsible glitch is large enough.  However, we again point out that our calculated GW amplitudes are in fact upper limits, where the limiting value assumes the energy of each glitch goes entirely into producing GWs (see Section~\ref{sec:gwglitch}).

\subsection{Scenario~2: NS and QS EoSs with somewhat overlapping $\nugw$
and $\tgw$} \label{sec:scenario2}
\begin{figure}
    \centering
    \includegraphics[width=0.48\textwidth]{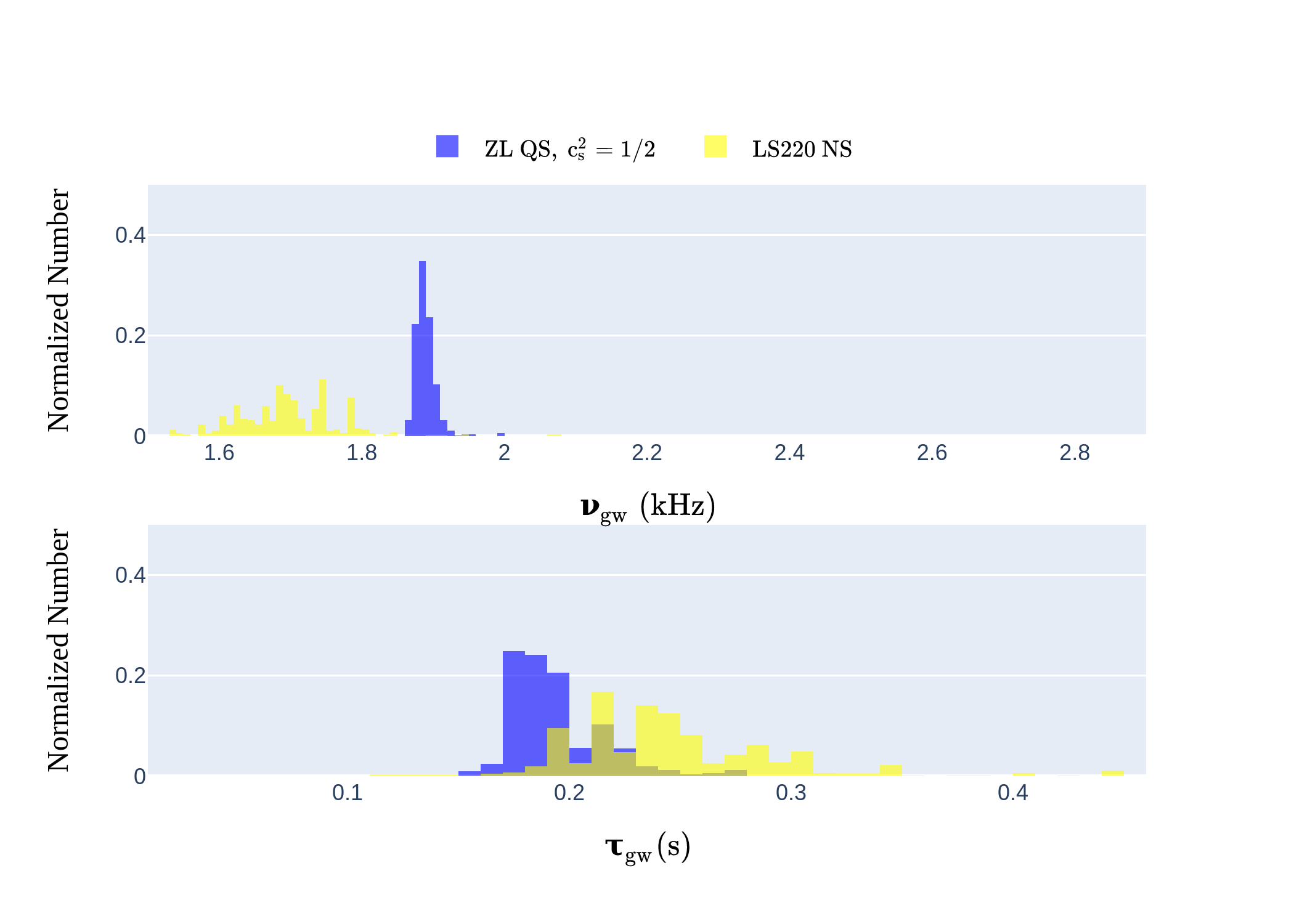}
\caption{Normalized distributions of GW frequency \(\nugw\) and damping time $\tgw$ using the ZL QS $\css=1/2$ EoS and LS220 NS EoS and Population 1 (known pulsars and glitches and assuming a Gaussian mass distribution).}
    \label{fig:ft2}
\end{figure}

\begin{figure}
    \centering
\includegraphics[width=0.48\textwidth]{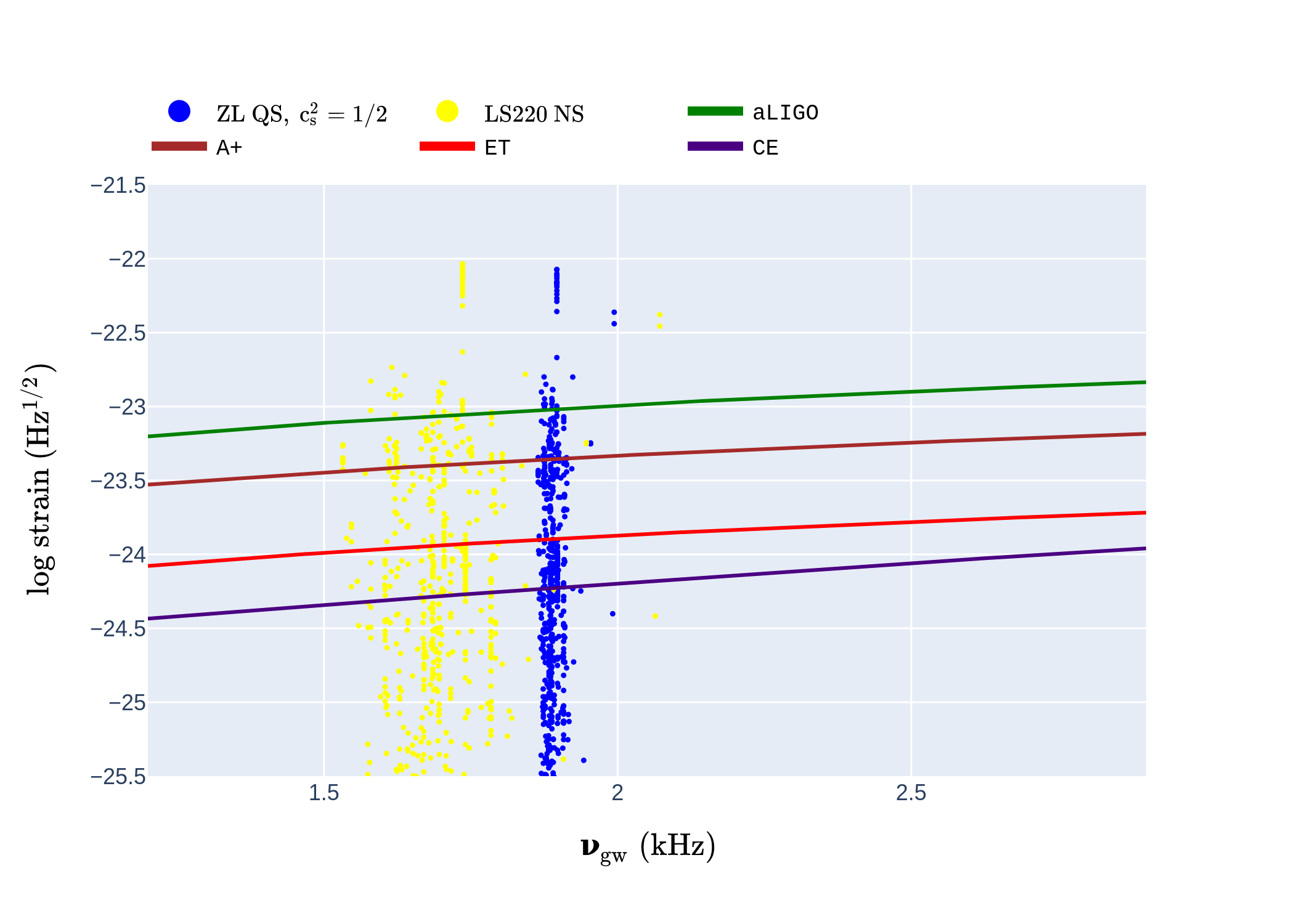}
\caption{GW strain ($h_0\sqrt{\tgw}$) from glitch-induced f-modes, calculated using the ZL QS $\css=1/2$ EoS and LS220 NS EoS and Population 1 (known pulsars and glitches and assuming a Gaussian mass distribution).  Solid lines indicate current and future GW detector sensitivities.}
    \label{fig:gw2}
\end{figure}

\begin{figure}
    \centering
\includegraphics[width=0.48\textwidth]{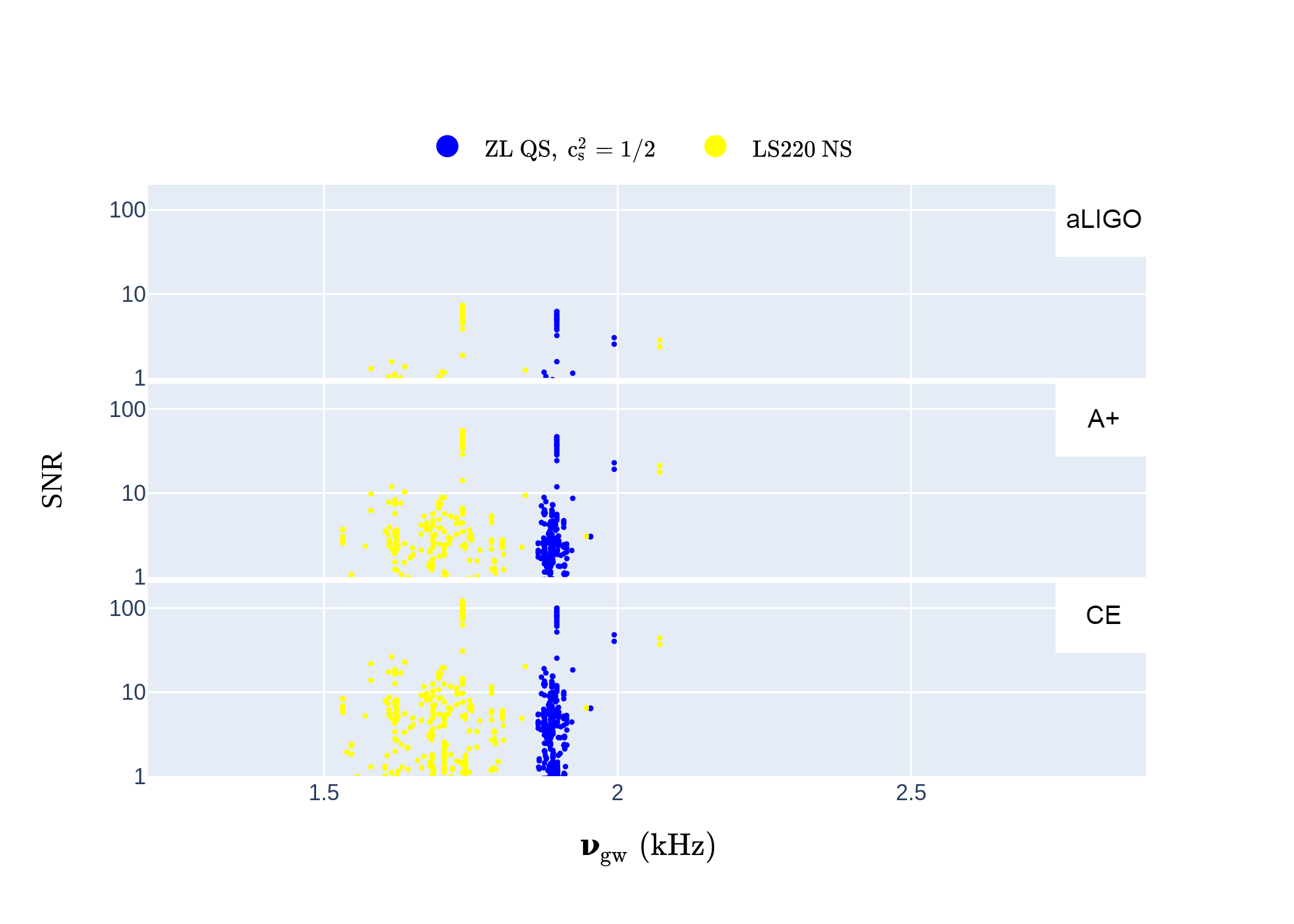}
\caption{Signal-to-noise ratio (SNR $=h_0\sqrt{\tgw/2S_h}$, where $\sqrt{S_h}$ is the spectral noise density of aLIGO, A+, or CE) from glitch-induced f-modes, calculated using the ZL QS $\css=1/2$ EoS and LS220 NS EoS and Population 1 (known pulsars and glitches and assuming a Gaussian mass distribution).} 
    \label{fig:snr2}
\end{figure}

We consider next the scenario when there is some overlap between the GW spectra of NSs and QSs.  We use results from the LS220 NS EoS and the ZL QS $\css=1/2$ EoS to illustrate this case.
Figures~\ref{fig:ft2}, \ref{fig:gw2}, and \ref{fig:snr2} show the GW frequency and damping time distributions, the GW spectra, and the SNRs, respectively.
We can see that the distributions for each EoS have some overlapping GW signals, but there is a significant difference between their respective means. Assuming a sufficiently even split between the number of actual QSs and NSs, it should be possible to determine that there are two groups in the data, particularly because QS EoSs produce such narrow distributions that QSs would appear as a highly concentrated group of sources away from the mean frequency for NSs.

\subsection{Scenario~3: NS and QS EoSs with significantly overlapping $\nugw$
and $\tgw$} \label{sec:scenario3}
\begin{figure}
    \centering
    \includegraphics[width=0.48\textwidth]{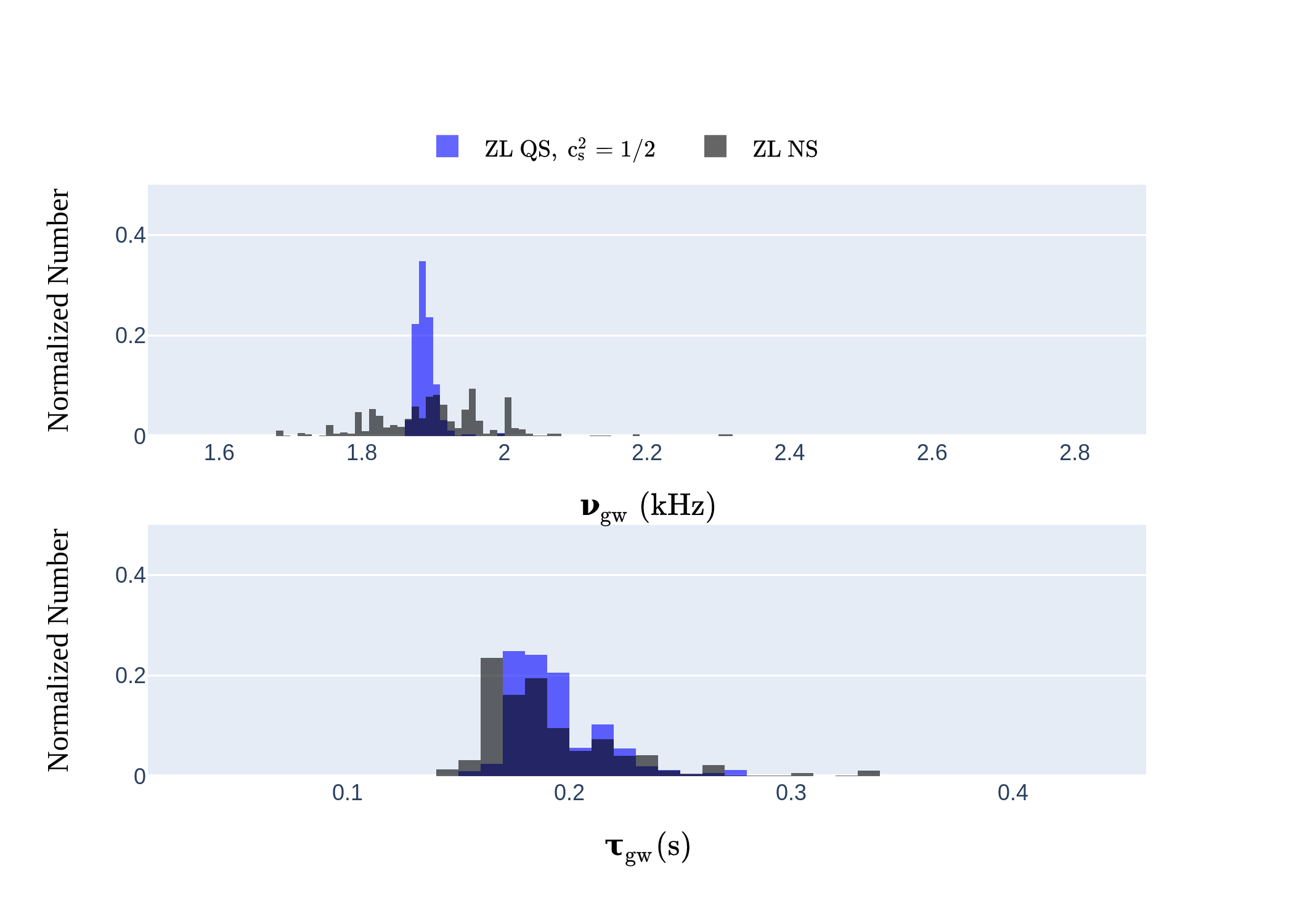}
\caption{Normalized distributions of GW frequency \(\nugw\) and damping time $\tgw$ using the ZL QS $\css=1/2$ EoS and ZL NS EoS and Population 1 (known pulsars and glitches and assuming a Gaussian mass distribution).}
    \label{fig:ft3}
\end{figure}

\begin{figure}
    \centering
\includegraphics[width=0.48\textwidth]{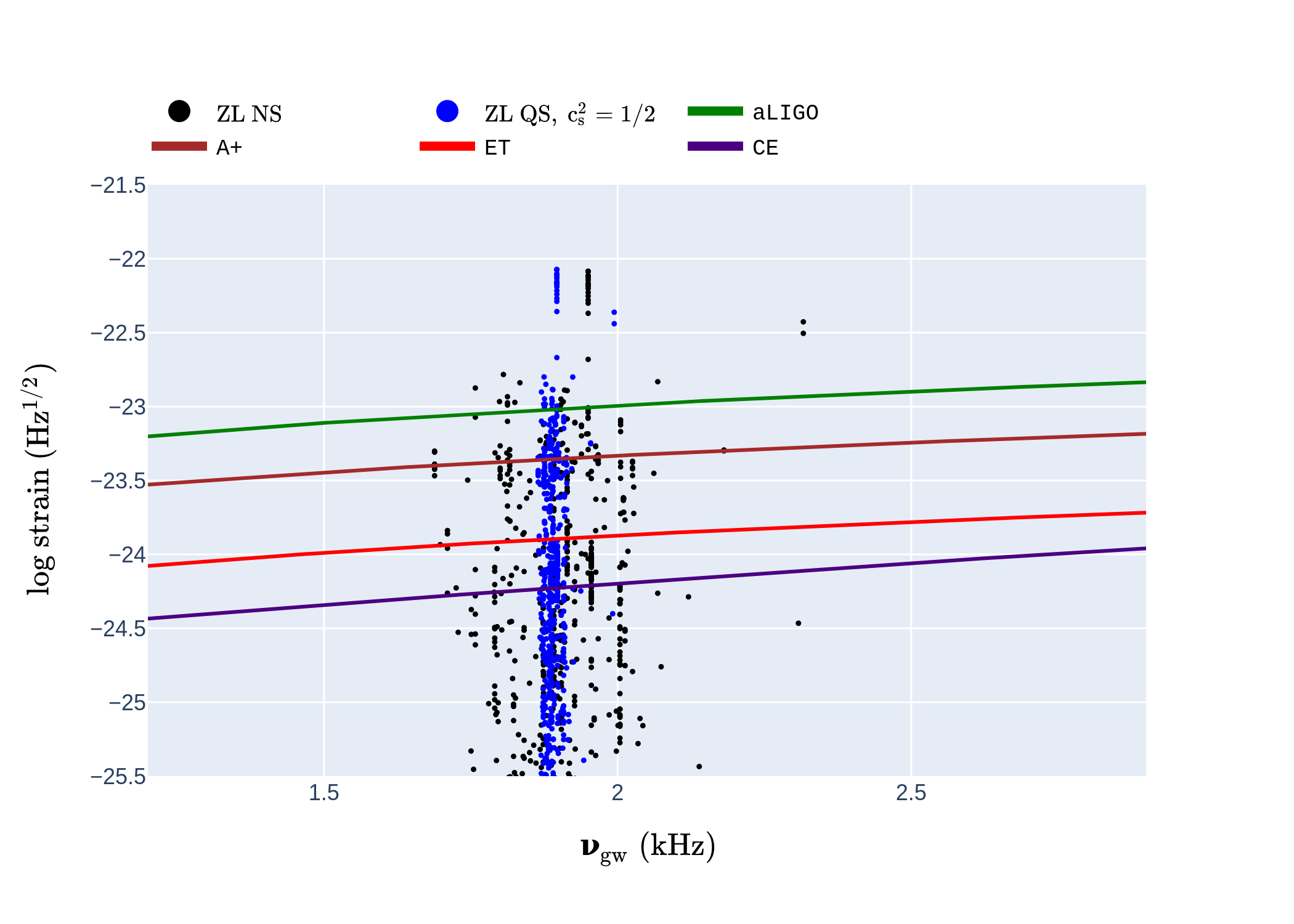}
\caption{GW strain ($h_0\sqrt{\tgw}$) from glitch-induced f-modes, calculated using the ZL QS $\css=1/2$ EoS and ZL NS EoS and Population 1 (known pulsars and glitches and assuming a Gaussian mass distribution).  Solid lines indicate current and future GW detector sensitivities.}
    \label{fig:gw3}
\end{figure}

\begin{figure}
    \centering
\includegraphics[width=0.48\textwidth]{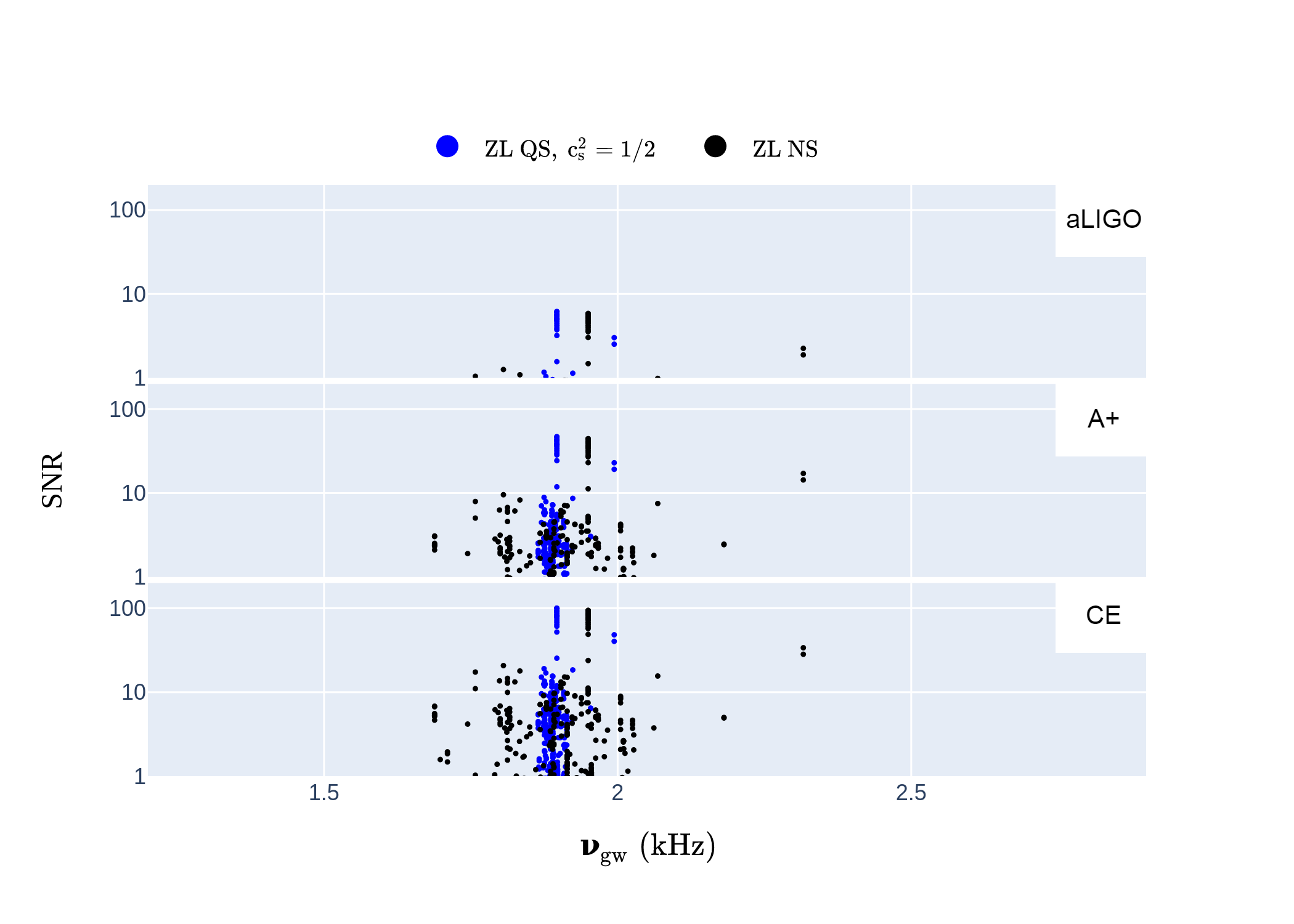}
\caption{Signal-to-noise ratio (SNR $=h_0\sqrt{\tgw/2S_h}$, where $\sqrt{S_h}$ is the spectral noise density of aLIGO, A+, or CE) from glitch-induced f-modes, calculated using the ZL QS $\css=1/2$ EoS and ZL NS EoS and Population 1 (known pulsars and glitches and assuming a Gaussian mass distribution).} 
    \label{fig:snr3}
\end{figure}

Finally, we consider the scenario when there is significant overlap between the GW spectra of NSs and QSs.
We use results from the ZL NS EoS and the ZL QS $\css=1/2$ EoS to illustrate this case.
Figures~\ref{fig:ft3}, \ref{fig:gw3}, and \ref{fig:snr3} show the GW frequency and damping time distributions, the GW spectra, and the SNRs, respectively.
In this scenario, the mean of the NS and QS frequency distributions are very similar, making it difficult to determine the existence of two groups of sources. If the QS fraction were high enough, it may be possible to determine the presence of QSs with enough data. However if it were as low as, for example, 10\%, it would be extremely difficult to distinguish between a concentration of sources resulting from the presence of a few QSs among many NSs.

\subsection{Some general detectability trends}
From the SNRs plotted in Figures~\ref{Fig. 8}, \ref{fig:snr2}, and \ref{fig:snr3}, we see that aLIGO does not seem to have sufficient sensitivity to achieve $\mbox{SNR}>1$ for very many known pulsars besides Vela, but the more advanced detectors do. Furthermore, the much more abundant number of sources from the simulated
Population 3 allows us to estimate the portion of Galactic spinning NS and QS glitches which would be ``in range'' (having an SNR of 1 or more) of each detector. Higher-frequency GWs are more difficult to detect,
so EoSs with lower-frequency f-modes cause a greater portion of glitches to be in range. Of the NS EoSs included here, the SHT EoS has the lowest f-mode frequencies and therefore the highest portion of in-range glitches, with 0.38\%, 1.5\%, 11\%, and 23\% being in range of the aLIGO, A+, ET, and CE observatories, respectively. In contrast, APR EoS has the highest f-mode frequencies, with only 0.11\%, 0.21\%, 1.7\%, and 4.6\% being in range. For QS EoSs, CMOT1 EoS with the CFL fit produces the lowest f-mode frequencies, causing 0.28\%, 0.89\%, 7.9\%, and 20\% to be in range, respectively, and ZL EoS with \(\css=1\) produces the highest f-mode frequencies, causing 0.11\%, 0.27\%, 2.0\%, and 5.7\% to be in range, respectively.

Based on the estimate in \cite{Faucher_Giguere_2006} that there are a total of $\sim 1.2\times10^6$ active radio pulsars (including non-observable rapidly spinning pulsars) in the Milky Way and that slightly fewer than 1 in 10 known pulsars have been observed to glitch (\(\sim\)200 of 3000), we can infer that somewhere on the order of $10^5$ glitching pulsars exist in the Galaxy. Using the above percentages for each GW detector, we can infer that an average glitch from about 100--400 pulsars would be in range for aLIGO, 200--2000 pulsars for A+, 2000--$1\times10^4$ pulsars for ET, and 5000--$2\times10^4$ pulsars for CE. It is worth noting that, particularly for the aLIGO sensitivities, these values may be overestimates. If we assume an average glitch period of 10 years, this implies detections of up to 40 glitches per year by aLIGO, 300 per year by A+, and detections every 6 hours and 4 hours on average for ET and CE respectively. The fact that current GW detectors have yet to detect a glitch at the time of writing suggests that these detection rates are overestimates, whether that be from the Sun's position, assumption of 100\% energy efficiency, or another reason. Nevertheless, these numbers illustrate the potential power of planned upgrades and new detectors, as well as provide an idea of how many undetected pulsars could be revealed through GW astronomy. It is also interesting to note that the EoS can affect the number of detectable glitches for a given detector by as much as an order of magnitude.
This means that if the NS and QS EoSs have sufficiently dissimilar f-mode frequencies, the ratio between the two observed source types could be skewed, either counter to or in addition to the actual imbalance in the split between the two populations.  With sufficient data, the bias could easily be corrected for, given that the effect of the sensitivity curves would be known. 

\section{Discussion}
GW astronomy is a rapidly expanding field that is being used to explore many objects and events in the Universe, including compact stars, supernovae, and the Big Bang. In addition to generating powerful GW signals during collisions, the matter in compact stars is dense enough that transient events within the star itself could have sufficiently high energies to produce detectable GWs. These GWs could be used to constrain the EoS for the matter in the star and could reveal that there is more than one type of EoS at play in the population, implying the existence of two types of  compact star. One of these two types could be QSs, composed entirely or almost entirely of stable quark matter. The existence of stable quark matter would essentially confirm the strange matter hypothesis, so the observation of GWs from pulsars is a way to explore this issue. 

One of the key findings of our study is that the QS EoSs that were used here produced f-mode frequency distributions that were universally more narrow than those produced by NS EoSs. This allowed us to examine three scenarios for the overlap of the potential NS and QS EoSs' GW spectra. Scenario 1 (distinct distributions of $\nugw$; see Section~\ref{sec:scenario1}) would present the least challenge in terms of determining the existence of QSs, as sufficient separation between the GW frequencies of just 2 sources could be enough to indicate the influence of 2 EoSs in the data.  Scenario 2 (overlapping distributions; see Section~\ref{sec:scenario2}) would require more sources to identify a high concentration away from the NS mean frequency but would also be relatively easy to confirm with enough sources. Scenario 3 (superimposed distributions; see Section~\ref{sec:scenario3}) would be the most difficult to identify the presence of QSs, especially if combined with a low fraction of QSs. But the search for QSs might be helped by exploring additional information in conjunction with GW frequency. For example, several sources having very similar frequencies but significantly (and regularly) different damping times might suggest the influence of 2 EoSs.
The process of determining if QSs are present would also be made easier by independently placing precise bounds on their masses. Unfortunately, determining the mass of non-binary pulsars is very difficult, and \cite{Ho_2020} actually explored the idea of GWs caused by pulsar glitches in part due to their potential as a method of determining pulsar masses. That being said, some methods for independently placing bounds on pulsar masses are being explored (e.g., \cite{Ho_2015,Montoli_2020}). 

 A natural explanation for the broader f-mode frequency $\nugw$ of NSs compared to QSs is that $\nugw$ is approximately $\propto\sqrt{M/R^3}$ (i.e., density; \cite{Andersson1998TowardsGW}) and $R$ is roughly constant for $M$ between $\sim1$ and $2\,M_\odot$ for many NS EoSs, so that $M/R^3$ varies by $\sim 30\%$. On the other hand, for QSs, $R$ changes significantly, increasing by $\sim 10\%$ over the same mass range. Thus $M/R^3$ is, in comparison, approximately constant since the mass and radius changes cancel.  Our work motivates more comprehensive studies to examine $\nugw$ in QSs and NSs in order to determine if this behavior is truly universal. $\tgw$ could also be studied further in both kinds of stars, but the EoSs explored in this paper do not show an easily discernible correlation between a $\tgw$ distribution and whether said distribution was from an NS or QS EoS. 

 Our results for $\nugw$ and $\tgw$ in this paper are representative of all f-mode frequencies and damping times (not just those caused by glitches) for QSs and NSs in the mass ranges and EoSs used, because $\nugw$ and $\tgw$ were calculated using only the randomly assigned masses, independent of the glitch data. Therefore, if the narrower width of the f-mode frequency distribution for QSs does universally hold across all EoSs, then 
 we might expect to see this difference in f-modes detected in other circumstances. As discussed in Section~\ref{sec:level1}, f-modes can occur during NS/QS births and in compact star collisions, as well as in magnetar activity. Given that births and collisions are more energetic events, these perhaps offer more likely possibilities for detections of f-modes, although they may be less frequent. 
 With enough f-mode data of any cause or method of detection, the trends and the three scenarios described in this paper should be applicable, and it should be possible to determine whether or not QSs are present in the data.

We also made predictions for the number of glitching pulsars that could be detected by various GW observatories at signal-to-noise ratios greater than 1, 
although these numbers are likely overestimates to some degree. This is due to a couple of caveats in our analysis, the first being the assumption of 100\% energy efficiency in the conversion of glitch energy to GWs, even though our analysis is largely irrespective of glitch physics details (see, e.g., \cite{Antonelli_2022,Antonopoulou_2022,Haskell_2024}). 
It is also important to note that GWs from f-modes alone will not be enough to rule out the possibility of QSs, because they could hide in the data as in Scenario 3 if they were rare enough or they glitch or oscillate differently than expected or not at all. However, a complete lack of evidence for QSs in GW data would make their existence seem less likely. As discussed, the relevance of the results in this paper is dependent on glitch-induced f-mode oscillations actually producing detectable GWs, which is not proven. The results for and discussion of QS EoSs are also implicitly dependent on QSs existing at all, which is the subject of debate in the field, and yet to be proven. A final caveat is that hybrid stars, composed of a core of quark matter surrounded by a large amount of hadronic matter, were not considered in this paper, and such stars might act as a middle ground between QSs and NSs, blurring the distinction between the two. However, many models for hybrid stars do not predict very significant deviations from NSs in the mass-radius relationship or f-mode characteristics, so this is not as problematic as it might initially seem \cite{Zhao_Lattimer_2022}. 
\begin{acknowledgments}
    The authors appreciate the efforts of Benjamin Shaw for maintaining the Jodrell Bank Glitch Catalogue. O.~H.~W. acknowledges the help of Elana Sewell-Grossman in the English language editing of the paper. O.~H.~W. acknowledges financial support from the Haverford Koshland Integrated Natural Sciences Center (KINSC).
\end{acknowledgments}
\appendix

\section{Full Versions of Figures \label{sec:appendix}}
\begin{figure*}
    \centering
\includegraphics[width=\textwidth]{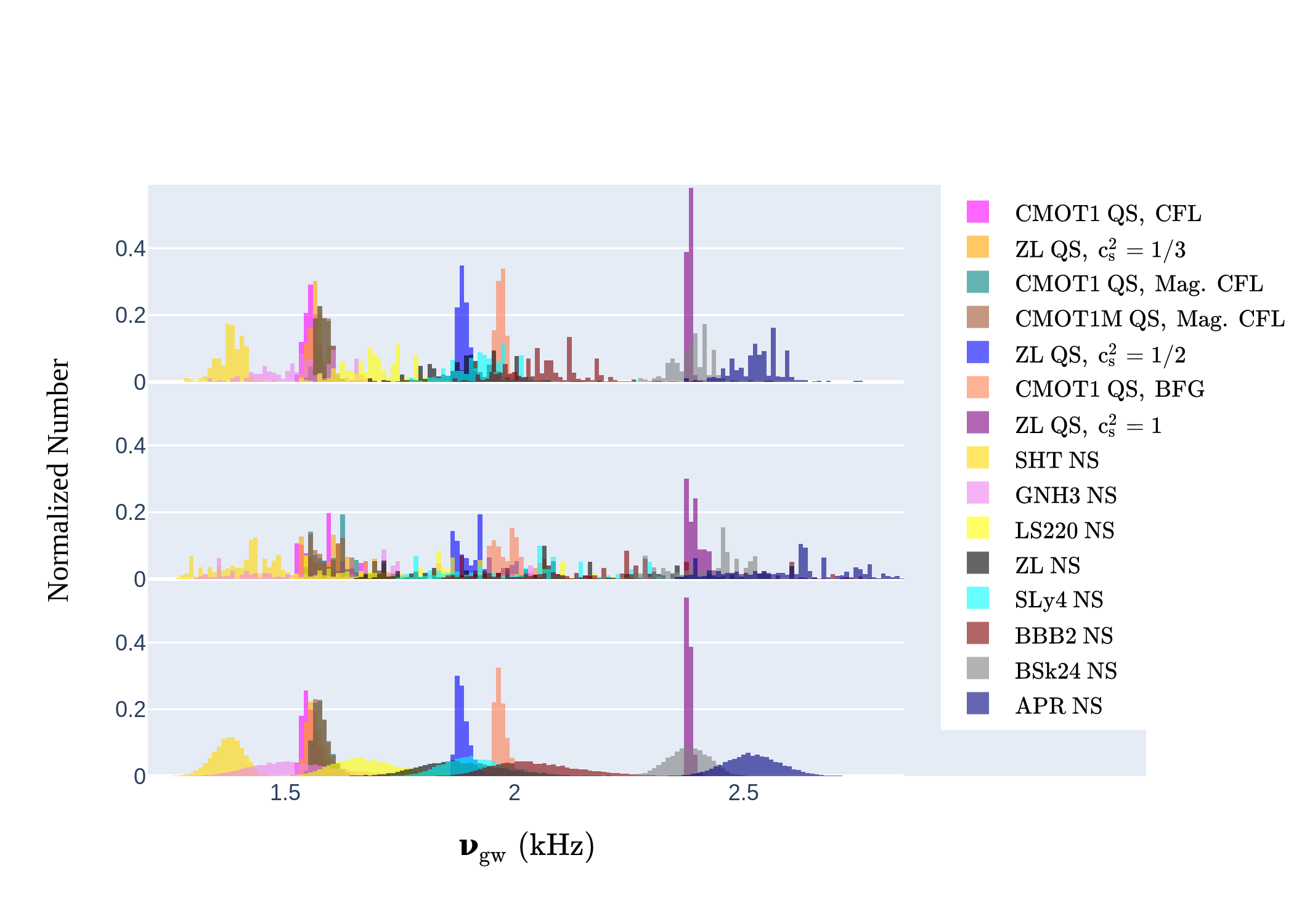}
\caption{Normalized distributions of GW frequency \(\nugw\) for all 15 NS and QS EoSs and Population~1 (top; known pulsars and glitches and Gaussian mass distribution), Population~2 (middle; known pulsars and glitches and uniform mass distribution), and Population~3 (bottom; $10^4$ simulated pulsars and glitches and Gaussian mass distribution).}
\label{Fig. 9}
\end{figure*}

\begin{figure*}
    \centering
\includegraphics[width=\textwidth]{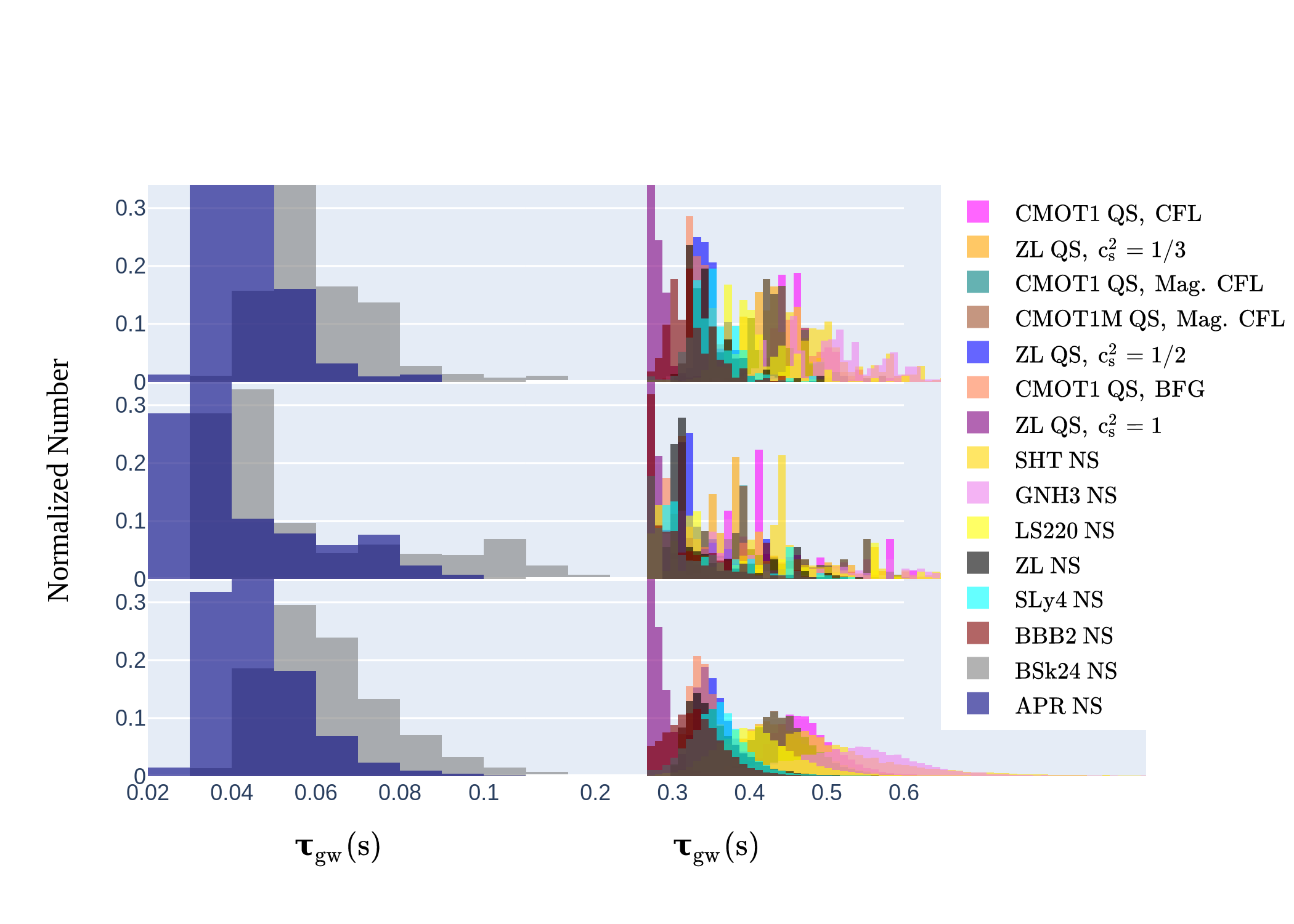}
\caption{Normalized distributions of GW damping time \(\tgw\) for all 15 NS and QS EoSs and Population~1 (top; known pulsars and glitches and Gaussian mass distribution), Population~2 (middle; known pulsars and glitches and uniform mass distribution), and Population~3 (bottom; $10^4$ simulated pulsars and glitches and Gaussian mass distribution).}
\label{Fig. 10}
\end{figure*}

\begin{figure*}
\includegraphics[width=\textwidth]{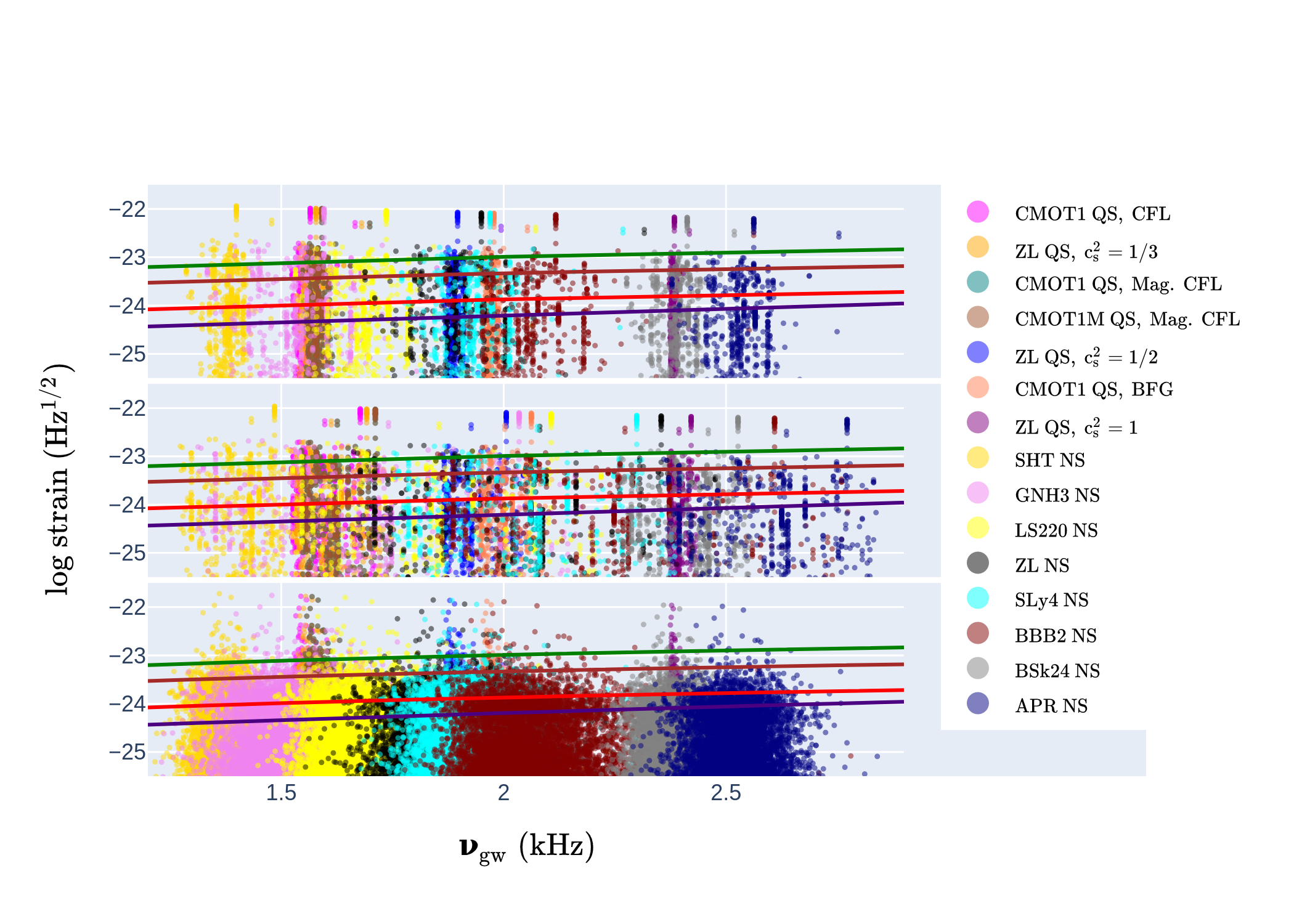}
\caption{GW strain ($h_0\sqrt{\tgw}$) from glitch-induced f-modes for all NS and QS EoSs and Population~1 (top panel; known pulsars and glitches and Gaussian mass distribution), Population~2 (middle panel; known pulsars and glitches and uniform mass distribution), and Population~3 (bottom panel; $10^4$ simulated pulsars and glitches and Gaussian mass distribution). Solid lines from top to bottom indicate GW detector sensitivities of aLIGO, A+, ET, and CE.}
\label{Fig11}
\end{figure*}

Figures~\ref{Fig. 9}--\ref{Fig11} are more complete versions of Figures~\ref{Fig. 5}, \ref{Fig. 6}, \ref{fig:ft2}, \ref{fig:gw2}, \ref{fig:ft3}, and \ref{fig:gw3} in that they show results for all 15 EoSs studied here compared to the 5 EoSs shown in the latter figures. CMOT1m refers to the magnetar version of the CMOT1 EoS. Note that the asymmetric shape of the distribution for the ZL QS \(\css=1\) EoS (purple) is because of the shape of the \(\tgw(M)\) function in \cite{Zhao_Lattimer_2022} for this EoS, where some masses are double-valued for certain damping times, which causes a large range of masses to produce very similar damping times. Also, the APR and BSk24 EoSs produce significantly smaller damping times than the other EoSs. As discussed in Section~\ref{sec:scenario1}, the trend of QSs having narrower and NSs having wider GW frequency distribution holds across all 15 EoSs, as can be seen here in Figures~\ref{Fig. 9} and \ref{Fig11}.

\bibliography{ms}

\end{document}